\documentclass[twoside]{dis09}
\usepackage[latin1]{inputenc}
\usepackage[dvips]{graphicx,epsfig,color}
\usepackage{wrapfig,rotating}
\usepackage{amssymb,amsmath,array}

\newcommand{\PO}{I\!\!P}
\newcommand{\RO}{I\!\!R}
\newcommand{\xpom}{x_{\PO}}

\begin{document}
\title{New developments in QCD analysis of inclusive diffraction 
at HERA}

\author{Laurent Schoeffel
%
\vspace{.3cm}\\
%
CEA Saclay, Irfu/SPP, 91191 Gif-sur-Yvette Cedex, France
}

\maketitle

\begin{abstract}
A large collection of results for the
 diffractive  dissociation of virtual photons, $\gamma^{\star}p \to Xp$, have been obtained 
with the H1 and ZEUS detectors at HERA.
Different experimental techniques have been used,
by requiring a large rapidity gap between 
$X$ and the outgoing proton, by analysing the mass 
distribution, $M_X$, of the hadronic final state, as well as by directly 
tagging the proton. A reasonable compatibility between those techniques and 
between H1 and ZEUS results have been observed.
Some common fundamental features
in the measurements are also present in all data sets. 
Diffractive PDFs can give a good account of those features. 
Ideas and new developments are discussed in these proceedings.
\end{abstract}

\section{Inclusive Diffraction at HERA}

\begin{figure}[tbp]
\begin{center}
\psfig{figure=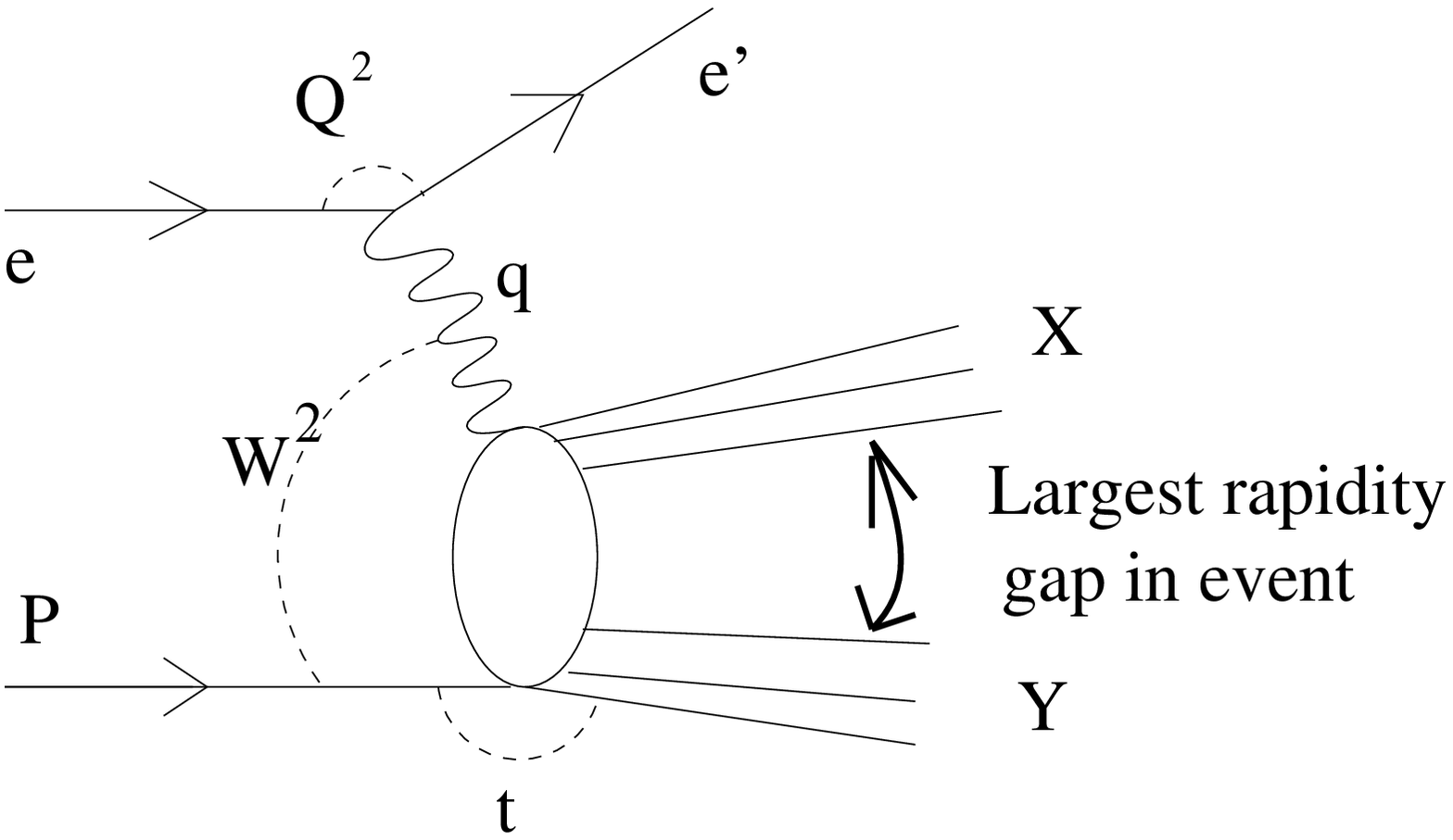,width=0.43\textwidth,angle=0}
\end{center}
\vspace{-1.cm}
\caption{Illustration of the process $ep \rightarrow eXY$. The
hadronic final state is composed of two distinct systems $X$ and
$Y$, which are separated by the largest interval 
in rapidity between final state hadrons.}
\label{difproc}
\end{figure}

One of the most important experimental result from the DESY $ep$ collider HERA
is the observation of a significant fraction of events in Deep Inelastic Scattering (DIS)
with a large rapidity gap (LRG) between the scattered proton, which remains intact,
and the rest of the final system. This fraction corresponds to about 10\% of the DIS   data
at $Q^2=10$ GeV$^2$.
In DIS, such events are not expected in such abundance, since large gaps are exponentially
suppressed due to color string formation between the proton remnant and the scattered partons.
Events
are of the type $ep \rightarrow eXp$, where the final state proton
carries more than $95$ \% of the proton beam energy. 
A photon of virtuality $Q^2$, coupled to the electron (or positron),
undergoes a strong interaction with the proton (or one of its 
low-mass excited states $Y$) to form a hadronic final state
system $X$ of mass $M_X$ separated by a LRG
from the leading proton (see Fig. \ref{difproc}). 
These events are called diffractive (see \cite{marta} for more details).
In such a reaction, $ep \rightarrow eXp$,
no net quantum number is exchanged and 
  the longitudinal momentum fraction $1-x_{\PO}$  
  is lost by the proton. Thus, the mongitudinal momentum $\xpom P$ is transfered 
to the system $X$. In addition to the standard DIS kinematic variables and $\xpom$, 
a diffractive event is also often 
characterised by the variable $\beta={x_{Bj}}/{x_{\PO}}$, which takes a simple 
interpretation in the parton model discussed in the following.

\begin{figure}[!]
\begin{center}
\includegraphics[width=8cm,height=6cm]{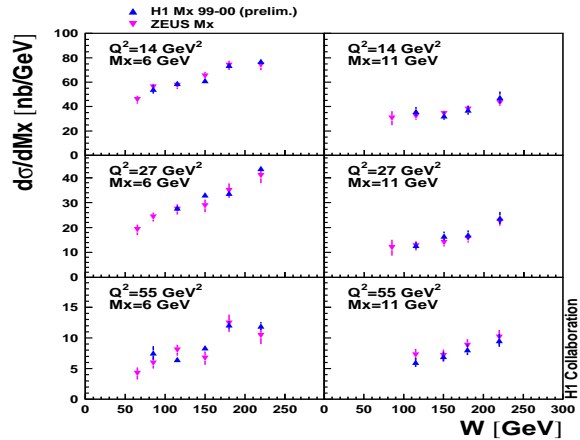}
\caption{Cross sections of the diffractive process $\gamma^* p \rightarrow p' X$, 
differential in the mass of the diffractively produced hadronic system $X$ ($M_X$),
are presented as a function of the centre-of-mass energy of the $\gamma^*p$ system $W$.
Measurements at different values of the virtuality
$Q^2$ of the exchanged photon are displayed. We observe a behaviour of the form $\sim W^{ 0.6}$  for the diffractive cross section, 
compatible with the dependence expected
for a hard process. 
}
\label{figdata}
\end{center}
\end{figure}

In order to compare diffractive data with perturbative QCD models, 
or parton-driven models,
the first step is to show that 
the diffractive cross section
shows a hard dependence in the centre-of-mass energy $W$ of the $\gamma^*p$ system.
In Fig. \ref{figdata}, we observe a behaviour of the form $\sim W^{ 0.6}$ , 
compatible with the dependence expected
for a hard process. This  observation is obviously 
the key to allow further studies of the diffractive process in the context of
perturbative QCD.
Events with the diffractive topology can be studied  in terms of
Pomeron trajectory exchanged between the proton and the virtual photon.
In this view, these events result from a colour-singlet exchange
between the diffractively dissociated virtual photon and the proton. 
Extensive measurements of diffractive DIS cross sections have been made by both
the ZEUS and H1 collaborations at HERA,
using different experimental techniques \cite{marta}. 
Of course, the comparison of these techniques provides a rich
source of information to get a better understanding of the experimental gains
and prejudices of those techniques.

\section{Diffractive PDFs}

A diffractive
structure function $F_2^{D(3)}$ 
can then be defined as a sum of two factorized 
contributions, corresponding to a Pomeron and secondary Reggeon trajectories: \\
$
F_2^{D(3)}(Q^2,\beta,x_{\PO})=
f_{\PO / p} (x_{\PO}) F_2^{D(\PO)} (Q^2,\beta) \
+ f_{\RO / p} (x_{\PO}) F_2^{D(\RO)} (Q^2,\beta)
$,
where $f_{\PO / p} (x_{\PO})$ is the Pomeron flux. It depends only on $\xpom$,
once integrated over $t$, and
$F_2^{D(\PO)}$ can be interpreted as the Pomeron structure function,
depending on $\beta$ and $Q^2$.
The other function,
$F_2^{D(\RO)}$, is an effective Reggeon structure function
taking into account various secondary Regge contributions which can not be 
separated.
The Pomeron and Reggeon fluxes are assumed to follow a Regge behaviour with  
linear
trajectories $\alpha_{\PO,\RO}(t)=\alpha_{\PO,\RO}(0)+\alpha^{'}_{\PO,\RO} t$, 
such that
\begin{equation}
f_{{\PO} / p,{\RO} / p} (x_{\PO})= \int^{t_{min}}_{t_{cut}} 
\frac{e^{B_{{\PO},{\RO}}t}}
{x_{\PO}^{2 \alpha_{{\PO},{\RO}}(t) -1}} {\rm d} t.
\label{flux}
\end{equation}

\begin{figure}[htbp]
\begin{center}
\includegraphics[totalheight=10cm]{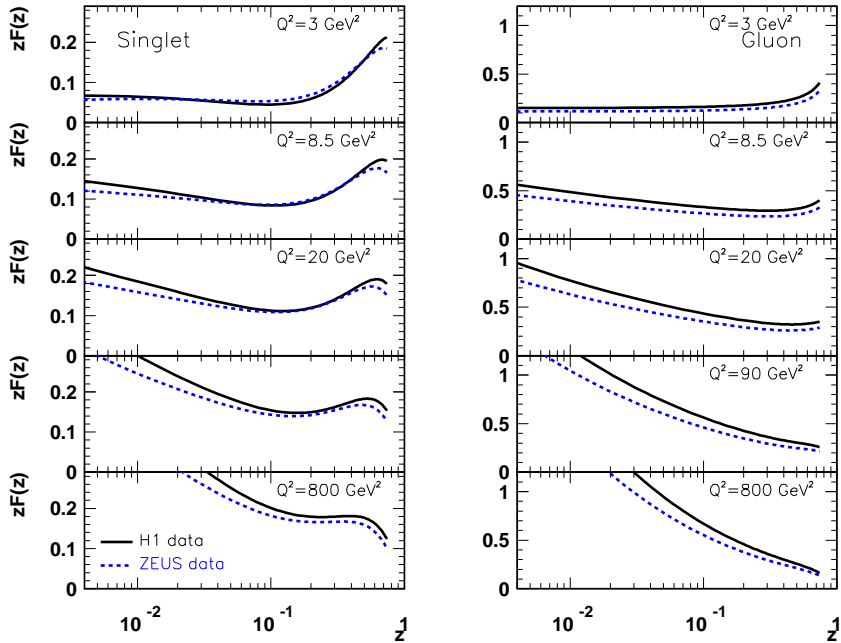}
\end{center}
\vspace{-0.5cm}
\caption{ Singlet and gluon distributions 
of the Pomeron as a function of $z$, the fractional momentum of the
Pomeron carried by the struck parton, derived from QCD fits on H1 and 
 ZEUS inclusive diffractive data (LRG)\cite{marta}. 
The parton densities are normalised to represent 
$\xpom$ times the true parton densities multiplied by the flux factor at
$\xpom = 0.003$ \cite{lolo1}. 
A good agreement is observed between both diffractive PDFs, which indicates that
the underlying QCD dynamics derived in both experiments is similar.
}
\label{figfin}
\end{figure}

Then, following \cite{lolo1}, diffractive PDFs can be extracted.
In Fig. \ref{figfin} we present the result for these diffractive PDFs
(quark singlet and gluon densities),
obtained using the most recent inclusive
diffractive cross sections presented in \cite{marta}.
In Fig. \ref{figfin}, the complete sets of published diffractive measurements
are included in the QCD fits for both experiments H1 and ZEUS.
Note also that in all QCD fits, we let the global relative normalisation
of the data set as a free parameter
(with respect to H1 LRG sample) \cite{lolo1}. 
The typical uncertainties for the diffractive PDFs in Fig. \ref{figfin}
ranges from 5\% to 10\% for the singlet density and from 10\% to
25\% for the gluon distribution, with 25\% at large $z$
(which corresponds to large $\beta$
for quarks) \cite{lolo1}.
Within these uncertainties, a nice agreement between diffractive PDFs
extracted from H1 and ZEUS data is obtained. This shows in an indirect way 
that data sets are well compatible and the underlying dynamics is
similar in all sets.

\begin{figure}[htbp]
\begin{center}
 \includegraphics[totalheight=8cm]{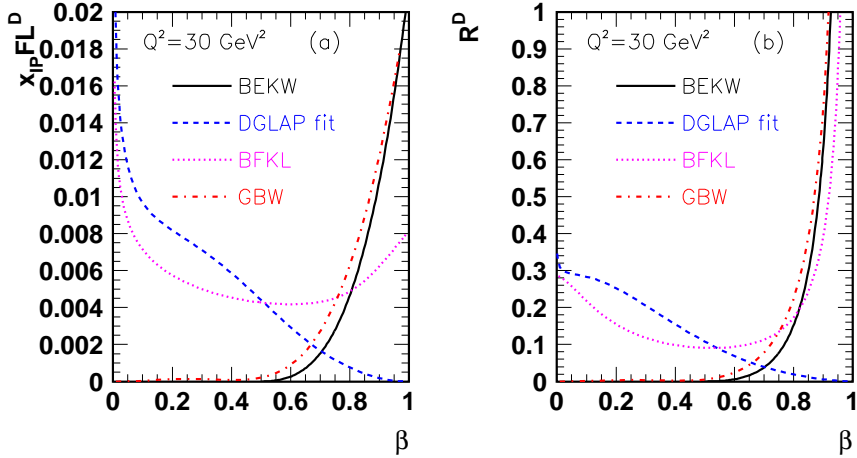}
\caption{ Predictions for $\xpom F_L^{D}$  and $R^D=\frac{F_L^D}{F_2^D-F_L^D}$  as a function of $\beta$
at $Q^2=30$~GeV$^2$ and $\xpom=10^{-3}$ \cite{lolo1}.
The dashed line prediction refers to the diffractive PDFs analysis
discussed in this part. 
Other curves
represent dipole model calculations.}
\label{Figdisc2}
\end{center}
\end{figure}

Then, an important conclusion  is the prediction for the
longitudinal diffractive structure function.
In Fig.~\ref{Figdisc2} we display this function with respect to
its dependence in $\beta$ 
 (Fig.~\ref{Figdisc2} (a)) and 
the ratio $R$ of the longitudinal to the transverse components 
of the diffractive structure function (Fig.~\ref{Figdisc2} (b)).

Note that diffractive distributions are process-independent
functions.  They appear not only in inclusive diffraction but also in
other processes where diffractive hard-scattering factorisation holds.  
The cross section of such a process can be
evaluated as the convolution of the relevant parton-level
cross section with the DPDFs.
For instance, the cross section
for charm production in diffractive DIS can be calculated at leading order
in $\alpha_s$ from the $\gamma^* g \rightarrow c \bar c$ cross section and
the diffractive gluon distribution.  An analogous statement holds for jet
production in diffractive DIS. Both processes have been analysed at
next-to-leading order in $\alpha_s$ and are found to be consistent with
the factorisation theorem.
A natural question to ask is whether one can use the DPDFs
extracted at HERA to describe hard diffractive processes such as the
production of jets, heavy quarks or weak gauge bosons in $p\bar{p}$
collisions at the Tevatron.  Using results on
diffractive dijet production from the CDF collaboration \cite{cdf}
compared to the expectations based on the 
DPDFs from HERA, a spectacular  discrepancy has been observed \cite{lolo1}.
Indeed, the fraction of diffractive dijet events at CDF is a factor 3 to 10
smaller than would be expected on the basis of the HERA data. The same
type of discrepancy is consistently observed in all hard diffractive
processes in $p\bar{p}$ events.  In
general, while at HERA hard diffraction contributes a fraction of order
10\% to the total cross section, it contributes only about 1\% at the
Tevatron.
This observation of QCD-factorisation breaking in hadron-hadron
scattering can be interpreted as a survival gap probability or a soft color interaction
which needs
to be considered in such reactions.
In fact, from a fundamental point of view, diffractive hard-scattering 
factorization does not apply to
hadron-hadron collisions.
Attempts to establish corresponding factorization theorems fail,
 because of interactions between spectator partons of the colliding
   hadrons.  The contribution of these interactions to the cross section
   does not decrease with the hard scale.  Since they are not associated
   with the hard-scattering subprocess, we no
   longer have factorization into a parton-level cross section and the
   parton densities of one of the colliding hadrons. These
interactions are generally soft, and we have at present to rely on
phenomenological models to quantify their effects \cite{cdf}. 
The yield of diffractive events in hadron-hadron collisions is then lowered
precisely because of these soft interactions between spectator partons
(often referred to as reinteractions or multiple scatterings).  
They can produce additional final-state particles which fill the would-be
rapidity gap (hence the often-used term rapidity gap survival).  When
such additional particles are produced, a very fast proton can no longer
appear in the final state because of energy conservation.  Diffractive
factorization breaking is thus intimately related to multiple scattering
in hadron-hadron collisions. Understanding and describing this
phenomenon is a challenge in the high-energy regime that will be reached
at the LHC \cite{afp}.

\section{Summary and outlook}

We have presented and discussed the most recent results on extraction of
diffractive PDFs from the HERA experiments.
A large collection of data sets and diffractive cross sections 
are published, which present common fundamental
features in all cases. 
The different experimental techniques, for both H1 and ZEUS
experiments, provide compatible results, with still some global normalisation
differences of about 10\%.
DPDFs give a good account of the main features of the all diffractive data sets.
This is a very important message from HERA that diffractive DPFs are well compatible
for both experiments.
Of course,
there is still much to do on the experimental side
with large statistics analyses, in order to obtain a better understanding
of  data and backgrounds. QCD fits of these data will provide an interesting
tool for the purpose of combining data. This is a major challenge for
HERA experiments in the next year.


\begin{footnotesize}



%

\end{footnotesize}


\end{document}